\documentclass[conference]{IEEEtran}
\pdfoutput=1
\usepackage{float}
\usepackage{amsfonts}
\usepackage{amsthm}
\usepackage{amsmath}
\usepackage{amssymb}
\usepackage{graphicx}
\usepackage{hyperref}
\usepackage{algorithm}
\usepackage{algpseudocode}
\floatstyle{ruled}
\newfloat{algorithm}{tbp}{loa}
\providecommand{\algorithmname}{Algorithm}
\floatname{algorithm}{\protect\algorithmname}

\DeclareMathOperator{\supp}{supp}

\newcommand{\Lav}{L_{\mathord{\rm av}}}
\newcommand{\GF}{\mathrm{GF}}

\newcommand{\starbox}{\mathbin{\setbox0\hbox{$\square$}
\setbox1\hbox to \wd0{\hss$\star$\hss}
\protect\rlap{\box0}{\protect\raisebox{0.25ex}{\box1}}
}}
\begin{document}
\title{A variant of list plus CRC concatenated polar code}
\author{\authorblockN{Gregory Bonik}
\authorblockA{Dept. of Mathematics\\
 University of Connecticut\\
 196 Auditorium Road, Unit 3009\\
Storrs, CT 06269-3009, USA\\
Email: grigory.bonik@uconn.edu}
\and
\authorblockN{Sergei Goreinov}
\authorblockA{Institute of Numerical\\
Mathematics, R.A.S.\\
Gubkina 8, 119333 Moscow, Russia\\
Email: sergei@inm.ras.ru}
\and
\authorblockN{Nickolai Zamarashkin}
\authorblockA{Institute of Numerical\\
Mathematics, R.A.S.\\
Gubkina 8, 119333 Moscow, Russia\\
Email: kolya@bach.inm.ras.ru}}
\maketitle
\begin{abstract}
A new family of codes based on polar codes, soft concatenation and list+CRC decoding is proposed.
Numerical experiments show the performance competitive with industry standards and Tal, Vardy approach.
\end{abstract}
\section{Introduction}
Polar codes were invented by E.~Ar\i kan in 2008. They are the first coding system possessing, on the theorem level,
the convergence to Shannon limit for code length $N\to\infty$, as well as fast encoding/decoding algorithms with complexity bound
$O(N \log_2 N)$. Thus polar codes are a significant theoretical result.

On the other hand, the performance of polar codes in their initial form presented by Ar\i kan, is considerably inferior,
for a fixed code length, to other coding systems used in various communication system
standards \cite{wimax}. In order to improve the polar code performance for moderate block lengths, there were made some
proposals based on the idea of combining polar and various other codes
\cite{bakshi,trif1,tv}. However only the system Polar + List + CRC
introduced in Tal, Vardy \cite{tv} seems to have the potential of outperforming the coding systems currently used in industry.
This system in our opinion has some drawbacks, an important one being the significant growth of decoder complexity
for large list lengths, e.g. $L=32$. Important theoretical complexity bound $O(LN\log N)$\cite{tv} does not adequately show
large overhead of this method which can be seen in practice. Thereby construction of coding schemes of significantly lower
complexity with comparable or better performance than those of \cite{tv} is an important problem. In this paper we propose
a method of construction of high-performance polar coding based schemes which possesses decoding complexity lower than
that of \cite{tv}. Our approach is a combination of soft concatenation \cite{bgz} and list decoding idea supplied with
some fast (CRC) check of candidates. The paper contains only the schemes themselves and some numerical experiments. Justification,
optimal choice of parameters and thorough comparison with other methods are left for future work.

The paper is organized as follows. Section 2 contains description of soft concatenation schemes \cite{bgz}.
Combination of these schemes with idea pursued in Tal, Vardy \cite{tv} is presented in section 3. Section 4 contains
the results of numerical experiments obtained for the new coding scheme.

\section{Soft concatenation scheme}
In this section we consider a method of performance improvement for polar codes \cite{bgz}
in which short classic error correcting codes are used together with polar codes.

Let $C_{1},C_{2},\ldots,C_{q}$ be a set of linear codes of equal length $M$.
Let $K_{i}$ be the number of information bits in the code $C_{i}$.
Let $V$ be some $M\times N$ matrix each of whose elements
is $0$ or $1$. Denote by $v_{ji}$ with $0\le i<N$ and $0\le j<M$
the elements of $V$, by $v^j$ its row $j$
and by $v_i$ its column $i$. For all $i=\overline{0,N-1}$
choose some integer $a_i$ in the range $1$ to $q$. We consider only such matrices $V$
whose columns $v_i$ are codewords of $C_{a_i}$, i.e.
\begin{equation}
v_{i}\in C_{a_{i}},\quad i=\overline{0,N-1}.\label{eq:constraint}
\end{equation}

Consider an arbitrary polar code of length $N$ and rate $1$, i.e. without redundancy,
with matrix generator $G\in\GF(2)^{N\times N}$.
Encode each row of $V$ with this polar code obtaining a new matrix
$X\in\GF(2)^{M\times N}$:
\begin{equation}
X=VG.\label{eq:mtx}
\end{equation}
If the matrix $X$ is ``reshaped'' into a row, one can consider the set of all such
possible rows subject to restriction (\ref{eq:constraint})
as a linear code of length  $M\cdot N$ and rate
\[
\frac{K}{MN}=\frac{1}{MN}\sum_{i=0}^{N-1}K_{a_{i}}.
\]
Thus obtained linear code we will call the \emph{soft concatenated polar code}.
Let $Y$ be the matrix received after the transmission of $X$ through the channel
and let $y^j$ be its row $j$. The decoder works by applying alternatively
the steps of successive cancellation
method for rows of $Y$ and maximum likelihood decoder for its columns.

In order to decode the column $v_0$, compute for each row of $Y$
independently the logarithmic likelihood ratios
\[
L(v_{j,0})=\ln\frac{\Pr\{y^{j}|v_{j,0}=0\}}{\Pr\{y^{j}|v_{j,0}=1\}},
\]
just like in the usual successive cancellation method.
Then the values $L(v_{j,0})$, $j=\overline{0,M-1}$ gathered in a vector $y$ are
given as input to ML-decoder for the code $C_{a_0}$. The most likely codeword
$w\in C_{a_{0}}$ produced on output is taken as an estimate of $v_0$.

Next we compute the estimate of $v_1$. Assuming $v_0$ already known,
again compute for each row independently the LLRs
\[
L(v_{j,1})=\frac{\Pr\{y^{j}|v_{j,1}=0;\, v_{j,0}\}}{\Pr\{y^{j}|v_{j,1}=1;\, v_{j,0}\}},
\]
concatenate the values $L(v_{j,1})$ into a vector $y$, which will be the input of
ML-decoder for the code $C_{a_1}$. The obtained codeword is taken as an estimate
of $v_1$. Next, assuming $v_0$ and $v_1$ exactly known, compute the estimate of $v_2$
etc.

Note that the polar codes are a special case of concatenated polar codes
for $M=1$, $q=2$ and $C_{1}=\{0\},$ $C_{2}=\{0,1\}$. In this case, 
bit $i$ is frozen if $a_{i}=1$, and it is information bit, if $a_{i}=2$.

Let $E_i$ be the error probability for estimation of column $i$
under the constraint that all previous columns were estimated error-free.
Write the upper bound for block error probability:
\begin{equation}
P_{E}\leq\sum_{i=0}^{N-1}E_{i}.\label{eq:upper-bound}
\end{equation}

Fix some symmetric channel $W$, set of codes $C_{1},\ldots,C_{q}$
of length $M$, polar code of length $N$ and rate $1$.
We require to construct a concatenated polar code of given rate $k/N$,
i.e. choose numbers
$a_{0},a_{1},\ldots,a_{N-1}$ such that
\begin{equation}
\sum_{i=0}^{N-1}K_{a_{i}}=K.\label{eq:sum-k-equals-k}
\end{equation}
We will choose these numbers so as to minimize the upper bound (\ref{eq:upper-bound}).
Denote by $E_{i}^{k}$ the error probability for estimation of the column
$v_{i}$ under the constraint that all previous columns were estimated error-free
and $a_{i}=k$. Note that $E_{i}^{k}$ does not depend on $a_{j}$ for all
$j\neq i$. For a concrete choice of $a_{0},a_{1},\ldots,a_{N-1}$
we can write the following upper bound for $P_{E}$,
\begin{equation}
P_{E}\leq\sum_{i=0}^{N-1}E_{i}^{a_{i}}.\label{eq:upper-bound-re}
\end{equation}
Assume for now that for all $i=\overline{0,N-1}$ and $k=\overline{1,q}$
we can compute $E_{i}^{k}$. In this case, the optimal values of $a_{0},a_{1},\ldots,a_{N-1}$,
can be determined using the dynamic programming method \cite{bgz}.

Since the channel is symmetric
and the code is linear, we assume the all-zero codeword is sent.
Suppose that the columns $v_{0},v_{1},\ldots,v_{i-1}$ have been estimated correctly
and the decoder is to estimate $v_i$. Next the ML-decoder for the code $C_k$
takes on input the vector
\[
\lambda=[L(v_{0,i}),L(v_{1,i}),L(v_{2,i}),\ldots,L(v_{M-1,i})].
\]
For convenience, introduce the notation $\lambda_{j}\equiv L(v_{j,i})$. The components
of $\lambda$ are i.i.d. random variables.
Their probability function (or pdf) $f_{i}$
can be computed approximately \cite{bgz}.
We can assume that the column $v_i$ is transmitted via some symmetric channel
with LLR distribution
$f_{i}$.
Thus the problem of computing $E_{i}^{k}$ is reduced to the estimation
of error probability for the ML-decoder on a channel
with given probability function $f_i$.
It is well-known that the ML-decoder minimizes the linear functional
\[
\phi(c)=\sum_{j=0}^{M-1}c_{j}\lambda_{j},
\]
where $c=[c_{0},c_{1},\ldots,c_{M-1}]$ runs over all codewords of the code
$C_{k}$. For the all-zero codeword the functional $\phi$ is zero.
Hence if the decoding error occurs, there necessarily exists some
codeword $c'$ such that $\phi(c')\leq0$. The last inequality can be rewritten
as the sum of $w_{H}(c')$ terms,
\[
\sum_{j\,\in\,\supp c'}\lambda_{j}\leq0.
\]
Some nonzero codeword $c'$ will be strictly more preferable than $0$
if $\phi(c')<0$ and in this case the decoder error will surely occur.
If $\phi(c')=0$, the decoder may choose the correct codeword among those
which zero the functional $\phi$. For simplicity assume that $\phi(c')=0$
also implies the decoder error.
Write the probability of the event that for a fixed $c'$
the inequality $\phi(c')\leq0$ holds as
\[
\Pr\left\{ \sum_{j\,\in\,\supp c'}\lambda_{j}\leq0\right\} .
\]
The sum consists of $w_{H}(c')$ i.i.d. random variables with the probability function
$f_{i}$, therefore the probability function of the sum is
\[
f_{i}^{\star w_{H}(c')}\equiv\underbrace{f_{i}\star f_{i}\star\ldots\star f_{i}}_{w_{H}(c')\mbox{ times}}
\]
It follows that the probability of the event
$\phi(c')\leq0$ depends only on the weight $w$ of the codeword $c'$ and it can be written as
\[
P(f,w)=\sum_{x\in\supp f^{\star w}:\: x\leq0}f^{\star w}(x).
\]
The main contribution in the error probability is made by codewords of minimal weight.
Let $d_{k}$ be the code distance of the code $C_{k}$, and let $m_{k}$ be the number of
different codewords of weight $d_{k}$ in the code $C_{k}$.
Then the probability $E_{i}^{k}$ may be estimated as
\begin{equation}
E_{i}^{k}\approx m_{k}\cdot P(f_{i},d_{k}).\label{eq:estimate}
\end{equation}
\section{New scheme description}\label{sec:newdesc}
Our proposal is the combination of soft concatenation schemes and Tal--Vardy method. We start from a special case
when the dimensions of the matrix $X$ (\ref{eq:mtx}) are powers of two, $N=2^n$, $M=2^m$.

Each column of the matrix $X$ will be connected with some polar code chosen in a way which will minimize the block error probability.
Besides, each column of the matrix $X$ can contain certain number of bits (e.g. $4$, $8$ or $16$) reserved for CRC.
The choice of exact number of those bits we leave as an open question.
In our experiments, we use CRC-$4$ check for every column. The decoding procedure for such code consists
in alternatively applying steps of successive cancellation for rows of the matrix $Y$ and steps
of List+SC+CRC for its columns. Therefore each column has also a corresponding list size as a parameter which
is used in decoding.
 It is quite obvious that channels corresponding to columns of $X$
 with ``better'' statistical properties should correspond to smaller list sizes.
It is not difficult to show that the decoding complexity will be
$$
O(NM (\Lav \log M + \log N))
\eqno(*)
$$
where $\Lav$ is the average list size.

One may think that the estimate $(*)$ is quite similar to $O\bigl(LNM\log(NM)\bigr)$, where $L$ is the maximum list size.
However due to channel polarization we have in general $\Lav< L$.
Besides, the new scheme possesses some natural parallelism. Indeed, column bits for each step of row-wise successive cancellation
may be processed simultaneously and  the decoder required for this operation works with codewords of significantly smaller length.
Using the parallel construction one can reduce the decoder complexity to $O(N (\log N + \Lav M \log M)$.

In a more general scheme, the columns of the matrix $X$ represent the codewords of a different code family.
In \cite{bgz} (see also section 2) we have used the codes of length $32$.
For each of those codes, the maximum likelihood (Viterbi) decoder was used. Alternatively,
one can use list Viterbi decoding, see e.g. \cite{listvit}. Using additionally CRC bits
we obtain a construction where a variety of codes is used as internal codes achieving thus better performance.
Theoretical study of these issues goes beyond the scope of this article.

We note in addition that a scheme containing list decoding at the first step is quite possible. However the complexity
of the resulting decoder generally has a factor $L^2$. Nevertheless the idea of LDPC codes as internal ones seems to us
very attractive.
\section{Numerical experiments}
Using the approach described in section~\ref{sec:newdesc}, we have constructed two codes with $N=16$,
$M=32$ and rate values $13/16$, $3/4$.
The average list size used was $\Lav=8$.
The experiments show that the new code performance is comparable to existing industry standards
as well as to the Tal--Vardy scheme.
This suggests that the improvement of the proposed technology will result in quite competitive
codes.

One can notice that for large SNRs the error rate of the proposed scheme becomes worse
than that of other tested codes. At present, we cannot say whether this effect is intrinsic
to the proposed scheme or it can be removed with increase of outer codes length or some
other optimization.
\begin{figure}[h!]
\begin{centering}
\includegraphics[scale=0.45]{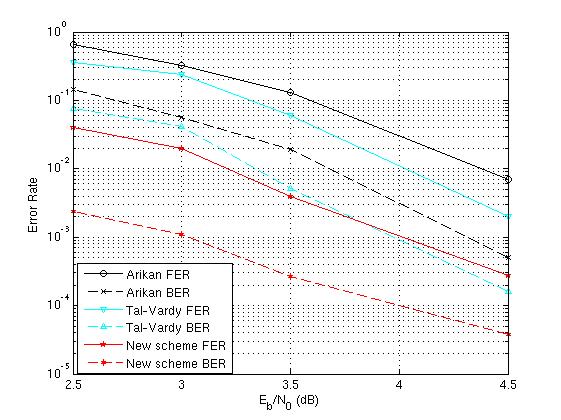}
\par\end{centering}
\caption{Comparison of performance of original Ar\i kan code, Tal--Vardy scheme ($L=8$) and proposed
scheme for code length $512$, rate $13/16$ on an AWGN channel\label{fig:htvn}}
\end{figure}
\begin{figure}[h!]
\begin{centering}
\includegraphics[scale=0.45]{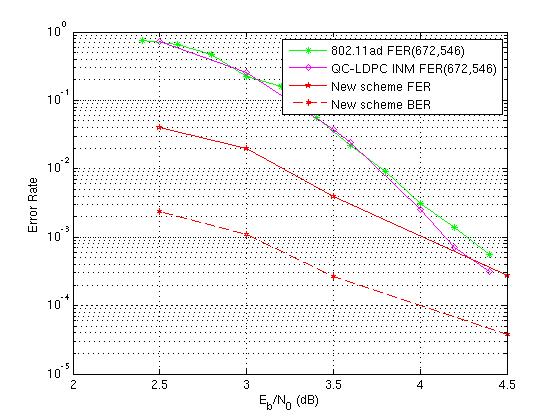}
\par\end{centering}
\caption{Comparison of performance of an LDPC code, 802.11ad standard and
proposed scheme on an AWGN channel. First two codes are $(672,546)$, while the third one is
$(512,416)$.\label{fig:hasn}}
\end{figure}
\begin{figure}[h!]
\begin{centering}
\includegraphics[scale=0.45]{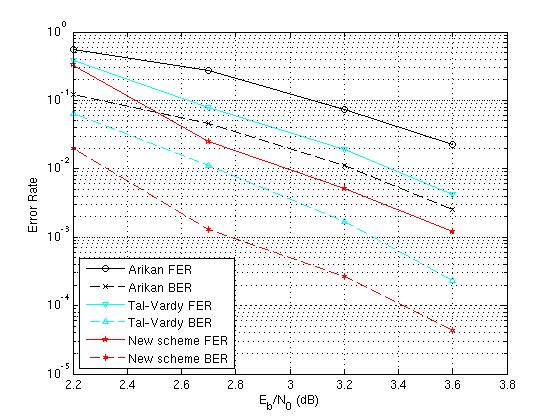}
\par\end{centering}
\caption{Comparison of performance of original Ar\i kan code, Tal--Vardy scheme ($L=8$) and proposed
scheme for code length $512$, rate $3/4$ on an AWGN channel.\label{fig:ltvn}}
\end{figure}
\begin{figure}[h!]
\begin{centering}
\includegraphics[scale=0.45]{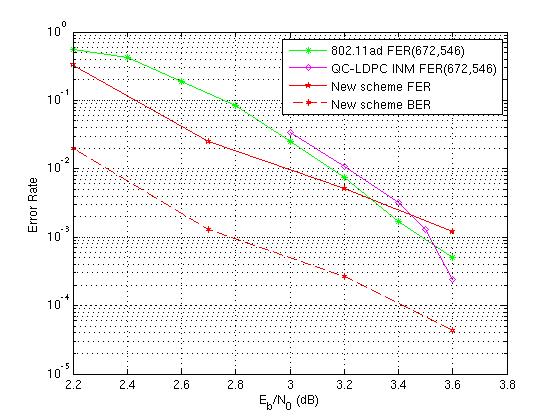}
\par\end{centering}
\caption{Comparison of performance of an LDPC code, 802.11ad standard and
proposed scheme on an AWGN channel.
First two codes are $(672,504)$ while the third one is $(512,384)$.\label{fig:lasn}}
\end{figure}


\begin{thebibliography}0
\bibitem{arik}
\sc E.~Ar\i kan,
\it Channel polarization: a method for constructing capacity-achieving codes
for symmetric binary-input memoryless channels,
\rm
IEEE Trans. Inform. Theory {\bf55}(7): 3051--3073 (2009).
\bibitem{bakshi}
\sc
M.~Bakshi, S.~Jaggi, M.~Effros,
\it
Concatenated Polar Codes,
\rm
http://arxiv.org/abs/1001.2545v1
\bibitem{trif1}
\sc P.~Trifonov, P.~Semenov,
\it
Generalized concatenated codes based on polar codes,
\rm
Proc.~8th Int. Symposium on Wireless Comm. Systems, ISWCS-2011, pp.~442--446.
\bibitem{tv}
\sc I.~Tal, A.~Vardy,
\it List Decoding of Polar Codes,
\rm http://arxiv.org/abs/1206.0050v1
\bibitem{wimax} IEEE 802.16e WiMax (Online) http://www.ieee802.org/16/tge/
\bibitem{bgz}
\sc G.~Bonik, S.~Goreinov, N.~Zamarashkin,
\it
Construction and analysis of polar and concatenated polar codes:
practical approach,
\rm http://arxiv.org/abs/1207.4343
\bibitem{listvit}
\sc
N.~Seshadri, C.-E.W.~Sundberg,
\it
List Viterbi decoding algorithms with applications,
\rm
IEEE Trans. on Comm. {\bf42}(234): 313--323 (1994).
\end{thebibliography}
\end{document}